
\input phyzzx
\hoffset=0.2truein
\voffset=0.1truein
\hsize=6truein
\def\TITLEPAGE{\frontpagetrue}
\def\CALT#1{\hbox to\hsize{\tenpoint \baselineskip=12pt
        \hfil\vtop{
        \hbox{\strut CALT-68-#1}}}}

\def\CALTECH{
        \address{California Institute of Technology,
Pasadena, CA 91125}}

\def\AUTHOR#1{\vskip .2in \centerline{#1}}
\def\ANDAUTHOR#1{\smallskip \centerline{\it and} \smallskip
\centerline{#1}}
\def\ABSTRACT#1{\vskip .2in \vfil \centerline{\twelvepoint
\bf Abstract}
        #1 \vfil}
\def\ENDTITLEPAGE{\vfil\eject\pageno=1}
\tolerance=10000
\TITLEPAGE
\CALT{1890}
\bigskip          
\titlestyle {Classical Electrodynamics with Dual Potentials}
\AUTHOR{M. Baker\foot{Work supported in part by the U.S. Dept. of Energy
under Contract No. DOE/ER/40614.}}
\centerline{\it University of Washington, Seattle, WA 98105}
\AUTHOR{James S. Ball}
\centerline{\it University of Utah, Salt Lake City, UT 84112}
\ANDAUTHOR{F. Zachariasen\foot{Work supported in part by the U.S. Dept. of
Energy
under Grant No. DE-FG03-92-ER40701.}}
\CALTECH
\ABSTRACT{We present Dirac's method for using dual potentials to solve
classical electrodynamics for an oppositely charged pair of particles,
with a view to extending these techniques to non-Abelian gauge theories.}

\ENDTITLEPAGE

\eject

\noindent {\bf I.   Solving Maxwell's Equations with Dual Potentials}

Dirac$^{[1]}$ showed that Maxwell's equations could be extended to include both
electrically and magnetically charged particles by connecting the magnetically
charged particles to strings.  In the absence of magnetically charged particles
one can apply Dirac's method to ordinary electrodynamics by connecting
electrically charged particles to strings.  In this formulation Maxwell's
equations become equations for dual potentials $C_\mu$ whose sources are the
polarization currents produced by the Dirac strings.  The potentials themselves
depend upon the location of the strings but they yield the same string
independent electromagnetic fields as the usual procedure.

 If in addition the dual potentials $C_\mu$ are minimally coupled to Higgs
fields, these fields necessarily carry magnetic charge.  Such a theory
describes the motion of electrically charged particles connected by Dirac
strings in a dual superconductor.  If extended to non-Abelian gauge
theory$^{[2]}$ it becomes a concrete realization of the Mandelstam
'tHooft$^{[3,4]}$  picture of color confinement as a manifestation of dual
superconductivity.  We have found that in order to understand this mechanism
for confinement it is very helpful to first have a clear picture of how dual
potentials work in ordinary electrodynamics.  Therefore, in this paper we
present an elementary discussion of Dirac's method applied to electrodynamics
and work out some simple examples.  Our discussion is entirely pedagogical and
contains nothing new although some specific results obtained here may not be
readily accessible elsewhere.  Our goal here is to make the following paper as
comprehensible as !
possible.

Consider first a sourceless linear dielectric medium.  Then Maxwell's equations
$$	\matrix{(a) & (b) & (c)\cr
\vec\nabla \cdot \vec D = 0 & \vec\nabla\cdot \vec B = 0 & \vec D = \epsilon
\vec E\cr
\vec\nabla \times \vec H - \partial_0 \vec D = 0 & ~~~\vec\nabla \times \vec E
+ \partial_0 \vec B = 0 & ~~~\vec B = \mu \vec H \,\, , \cr} \eqno (1.1)$$
can be solved by introducing vector potentials in either of two ways.  The
conventional choice is to write
$$	\vec B \equiv \vec\nabla \times \vec A, \quad \vec E \equiv - \partial_0
\vec A - \vec \nabla A_0, \eqno (1.2a)$$
in which case eqs. (1.1b) become kinematical identities and the dynamics is
contained in eqs. (1.1a).  The vector $A^\mu = (A_0, \vec A)$ is called the
vector potential.  The alternate (dual) choice is to write
$$	\vec D = - \vec \nabla \times \vec C, \quad \vec H = - \partial_0 \vec C -
\vec \nabla C_0, \eqno (1.2b)$$
in which case eqs. (1.1a) are kinematical identities and eqs. (1.1b) contain
the dynamics.  The vector $C^\mu = (C_0, \vec C)$ is called the dual vector
potential.

Let us first use $C_\mu$ to solve the source free Maxwell eqs. (1.1) in order
to get accustomed to using the dual potential.  We first write eqs.  (1.2b) in
covariant form by defining:
$$	G_{0k} = H_k, \quad G_{ij} = \epsilon_{ijk} D_k, \eqno (1.3)$$
so that eqs. (1.2b) take the form
$$	G_{\mu\nu} = \partial_\mu C_\nu - \partial_\nu C_\mu. \eqno (1.4)$$
In a relativistic medium $\epsilon = {1\over \mu}$.  Then using eq. (1.3) we
can write the constitutive equations as
$$	E_i = {\mu\over 2} \epsilon_{ijk} G_{jk}, \quad B_i =  \mu G_{0i}, \eqno
(1.5)$$
and Maxwell's eqs. (1.1b) as
$$	\partial^\alpha \mu G_{\alpha \beta} = 0. \eqno (1.6)$$
Eqs. (1.6) for $C_\mu$ have the same form as the usual Maxwell equations for
$A_\mu$, obtained from (1.1a), with the replacement $\mu \rightarrow \epsilon$,
and they are solved in the same way.  Eq. (1.5) then gives the electromagnetic
fields $\vec E$ and $\vec B$ in terms of $C_\mu$.

Electric current sources $j_\mu = (\rho, \vec j)$ appear only in eqs. (1.1a)
and not in eqs. (1.1b).  Hence in the presence of electric currents eqs. (1.1b)
remain valid and are still kinematic identities in terms of $A^\mu$.  Eqs.
(1.1a), in contrast, are no longer identities in terms of $C^\mu$.  However,
Dirac has shown how to generalize eqs. (1.2b) in order to satisfy eqs. (1.1a)
with dual potentials $C^\mu$ even in the presence of electric currents.

When charged particles are present eqs. (1.1a) become
$$	\matrix{ {\rm Gauss' ~Law} & {~~~~\rm Ampere's ~Law}\cr
\vec \nabla \cdot \vec D = \rho, & ~~~~~~~~\vec \nabla \times \vec H = \vec j +
{\partial \vec D\over \partial t}.\cr} \eqno (1.7)$$
Suppose that the total charge $Q = \int \rho d \vec x = 0$.  (If $Q \not= 0$,
then there will be Dirac strings extending to infinity, but nothing essential
will be changed.)  Then we can always find a polarization vector $\vec P$ and a
magnetization vector $\vec M$ so that
$$	\rho = - \vec \nabla \cdot \vec P, \quad \vec j = \vec \nabla \times \vec M
+ {\partial \vec P\over\partial t}, \eqno (1.8)$$
where $\vec P$ is the dipole moment per unit volume and $\vec M$ is the
magnetic moment per unit volume.  Inserting eq. (1.8) into (1.7), we obtain
$$	\vec\nabla \cdot (\vec D + \vec P) = 0, ~~\vec\nabla \times (\vec H - \vec
M) - {\partial (\vec D + \vec P)\over\partial t} = 0.$$
Hence,
$$	\vec D = - \vec\nabla \times \vec C - \vec P, \quad \vec H = - \vec \nabla
C_0 - {\partial\vec C\over\partial t} + \vec M. \eqno (1.9)$$
Eqs. (1.7) then become kinematical identities and eqs. (1.1b) contain the
dynamics as before. Using the definitions (1.3) of $G_{\mu\nu}$, we can write
eqs. (1.9) in the covariant form
$$	G_{\mu\nu} = \partial_\mu C_\nu - \partial_\nu C_\mu + G_{\mu\nu}^s, \eqno
(1.10)$$
where the tensor $G_{\mu\nu}^s$ has components
$$	G_{0k}^s = M_k, \quad G_{ij}^s = - \epsilon_{ijk} P_k, \eqno (1.11)$$
and we have now specialized to the case where  $\vec P$ and $\vec M$ arise from
Dirac strings connecting the charged particles;  hence the superscript ``s'' on
$G_{\mu\nu}^s$.   Eq. (1.10) is just the generalization of eq. (1.4) to account
for the presence of charged particles.  Eq. (1.11) shows that $G_{\mu\nu}^s$ is
 the dual of the polarization tensor.  Eqs. (1.1b) and (1.1c) are unchanged so
that eqs. (1.5) and (1.6) remain the same as does the definition (1.3) of
$G_{\mu\nu}$.  The effect of the charged particles is to change the relation
(1.4) between $G_{\mu\nu}$ and $C_\mu$ to (1.10) where  $G_{\mu\nu}^s$ is
determined in terms of $\rho$ and $\vec j$ by solving eqs. (1.8) for $\vec P$
and $\vec M$.  This is done in Section II for a pair of oppositely charged
particles.

Substituting eq. (1.10) into eq. (1.6) we obtain
$$ 	\partial^\alpha \mu (\partial_\alpha C_\beta -  \partial_\beta C_\alpha) =
- \partial^\alpha \mu G_{\alpha \beta}^s, \eqno (1.12)$$
which determines the dual potentials $C_\mu$ in terms of $G_{\mu\nu}^s$.  Eqs.
(1.12) provide an alternate form of Maxwell's equations which are completely
equivalent to the usual form expressed in terms of the vector potential
$A_\mu$, namely
$$	\partial^\alpha \epsilon (\partial_\alpha A_\beta - \partial_\beta A_\alpha)
= j_\beta. \eqno (1.13)$$
All text books on electricity and magnetism could be rewritten using only dual
potentials $C_\mu$ satisfying eq. (1.12) and the same electromagnetic forces
between charged particles would be obtained.  The potentials themselves,
however, could be completely different.  For example in a dielectric medium
having a wave number dependent dielectric constant $\epsilon (q) \rightarrow 0$
as $q^2 \rightarrow 0$ (corresponding to antiscreening at large distances), the
potentials $A_\mu$ determined from eq. (1.13) would be singular at large
distances, while the dual potentials $C_\mu$ satisfying eq. (1.12) with $\mu =
1/\epsilon \rightarrow \infty$ as $q^2 \rightarrow 0$ would be screened at
large distances.  Use of the potentials $A_\mu$ to describe this system would
introduce singularities which do not appear in the dual potentials $C_\mu$.
Hence the dual potentials are the natural choice to describe a medium with long
range antiscreening.

Note that for $\mu = \epsilon = 1, \vec B = \vec H, \vec D = \vec E$ and
substituting eqs. (1.9) in (1.1b) gives the equation for the dual potentials in
three-dimensional notation:
$$	\vec \nabla \cdot (-\vec \nabla C_0 - \partial_0 \vec C) = - \vec \nabla
\cdot \vec M, \eqno (1.14)$$
$$	\vec\nabla \times (- \vec\nabla \times \vec C) + \partial_0 (- \partial_0
\vec C - \vec\nabla C_0) = \vec \nabla \times \vec P - {\partial \vec
M\over\partial t}. \eqno (1.15)$$
These equations are identical to eq. (1.12) with $\mu = 1$.  They have the same
form as the equations for $A_\mu$, the ordinary vector potentials in a
polarizable medium with $\vec P$ and $\vec M$ interchanged.  For example $-
\vec\nabla \cdot \vec M$ is the source of $C_0$.  However eqs. (1.14) and
(1.15) describe the electrodynamics of electrically charged particles moving in
the vacuum and $\vec P$ and $\vec M$ are the polarization and magnetization
respectively of the Dirac strings attached to these particles, as we shall now
see.

\noindent {\bf II.  The Fields of a Pair of Oppositely Charged Particles}

We now apply the results of the previous section to the case of two particles
of charge $e (-e)$ moving along trajectories $\vec x_1 (t) (\vec x_2 (t))$ in
free space with $\mu = \epsilon = 1$. Then
$$	\rho (\vec x, t) = e [\delta^3 (\vec x - \vec x_1 (t)) - \delta^3 (\vec  x -
\vec x_2 (t))] , \eqno (2.1)$$
and
$$	\vec j (\vec x, t) = e [\vec v_1 \delta^3 (\vec x - \vec x_1 (t)) - \vec v_2
\delta^3(\vec x - \vec x_2 (t))], \eqno (2.2)$$
where  $\vec v_i = {d \vec x_i\over dt} , ~ i = 1, 2$.  We must find a
polarization $\vec P$ and magnetization $\vec M$ satisfying eq. (1.8) with
$\rho$ and $\vec j$ given by eqs. (2.1) and (2.2).  The solution of this
problem was given by Dirac.$^{[1]}$  Let $\vec y (\sigma, t)$ be any line
$L(t)$ connecting $\vec x_2 (t)$ and $\vec x_1(t)$ i.e., $\vec y (\sigma_1, t)
= \vec x_1(t), ~~ \vec y (\sigma_2,  t) = \vec x_2 (t),  \sigma_2 \leq \sigma
\leq \sigma_1$.  (See Figure 1.)
On each element $d\vec y$ of $L$  place a dipole moment $d\vec p = e d\vec y$.
It is evident from Fig. 1 that the charge and current density produced by the
sum of these dipoles is that due to the pair of moving oppositely charged
particles, namely eqs. (2.1) and (2.2).  To obtain (2.1)  formally we note that
the dipole moment per unit volume $\vec P$ is
$$	\vec P (x) = e \int_{\vec x_{2}(t)}^{\vec x_{1} (t)} d\vec y \delta (\vec x
- \vec y) = e \int_{\sigma_{2}}^{\sigma_{1}} d\sigma {\partial \vec y (\sigma,
t)\over \partial \sigma} \delta (\vec x - \vec y (\sigma, t)). \eqno (2.3)$$
Then
$$	- \vec\nabla \cdot \vec P = - e \int_{\vec x_{2}(t)}^{\vec x_{1}(t)} d\vec y
\cdot \vec \nabla_x \delta (\vec x - \vec y) = \rho (\vec x) . \eqno (2.4)$$
Furthermore since the line element $d\vec y$ is moving with velocity $\vec v =
{\partial\over\partial t} \vec y (\sigma, t)$, the string $L$ in Fig. 1 has a
magnetization
$$	\vec M = e \int_{\vec x_{2}}^{\vec x_{1}} d\vec y  \times {\partial \vec
y\over\partial t} \delta (\vec x - \vec y) = e \int_{\sigma_{2}}^{\sigma_{1}}
d\sigma {\partial\vec y\over\partial\sigma} \times {\partial\vec y\over\partial
t} \delta (\vec x - \vec y (\sigma,t)). \eqno (2.5)$$

Next we show explicitly that eq. (2.3) for $\vec P$ and (2.5) for $\vec M$ give
via eq. (1.8) the current density.  From eq. (2.3) we have
$$	{\partial \vec P\over \partial t} = e {\partial\over\partial t} \int_{\vec
x_{2}(t)}^{\vec x_{1}(t)} d\vec y (\sigma, t) \delta (\vec x - \vec y (\sigma,
t))$$
$$	= e \left[{d \vec x_1\over dt} \delta (\vec x - \vec x_1 (t)) - {d\vec
x_2\over dt} \delta (\vec x - \vec x_2 (t))\right]$$
$$	+ {e\over\delta t} \left[\int_{\vec x_{2}(t)}^{\vec x_{1}(t)} d\vec y
(\sigma, t + \delta t) \delta (\vec x - \vec y (\sigma, t + \delta t)) -
\int_{\vec x_{2}}^{\vec x_{1}} d\vec y(\sigma, t) \delta (\vec x - \vec y
(\sigma, t))\right]$$
$$	= \vec j (\vec x,  t) + {e\over\delta t} \oint d\vec y \delta (\vec x - \vec
y). \eqno (2.6)$$
The first term on the right-hand side of eq. (2.6) arises from differentiating
with respect to the end points with the path fixed.  The line integral in eq.
(2.6) is over a closed contour running from $\vec x_2(t)$ to $\vec x_1(t)$
along the path $\vec y (\sigma, t +\delta t)$ and returning to $\vec x_2(t)$
along $\vec y (\sigma, t)$.   (See Fig. 2.)
We denote $\vec y (\sigma, t +  dt) - \vec y (\sigma, t) = \delta \vec y$ and
the element of area $d\vec y \times \delta \vec y \equiv d\vec S$.  Then
by Stokes' theorem
$$	{e\over \delta t} \oint d \vec y \delta (\vec x - \vec y) = - {e\over \delta
t} \int d\vec S \times  \vec\nabla_y \delta (\vec x - \vec y)$$
$$	= e \int d\vec y \times {\delta \vec y\over\delta t} \times \vec\nabla_x
\delta (\vec x - \vec y) = - \vec\nabla \times \vec M . \eqno (2.7)$$
Eqs. (2.6) and (2.7)  yield eq. (1.8) with $\vec j$ given by eq. (2.2) as
asserted.

Eqs. (1.14) and (1.15) with $\vec P$ and $\vec M$ given by eqs. (2.3) and (2.5)
respectively determine $C_\mu$.  To obtain the explicit form of the covariant
version of these equations we note that

$$	G_{\mu\nu}^s = - e \epsilon_{\mu\nu\alpha\beta} \int_{\tau_{2}}^{\tau_{1}}
d\tau \int_{\sigma_{2}}^{\sigma_{1}} d\sigma {\partial y^\alpha\over
\partial\tau} {\partial y^\beta\over\partial\sigma} \delta^4 (x - y), \eqno
(2.8)$$
where
$$	x^\mu = (t, \vec x), \quad y^\mu = (y^0, \vec y),$$
and
$$	d\tau = \sqrt{(dy^0)^2 - (d\vec y)^2}.$$
Eq. (2.8) is the standard covariant  form for the Dirac string field
$G_{\mu\nu}^s$.$^{[1]}$  To show that the expressions (1.11) and (2.8) for
$G_{\mu\nu}^s$ are the same first set $\mu = 0$ and $\nu = k$ in eq. (2.8):
$$	G_{0k}^s = - e \epsilon_{kmn} \int_{\tau_{2}}^{\tau_{1}} d\tau
\int_{\sigma_{2}}^{\sigma_{1}} d\sigma {\partial y^m\over\partial\tau}
{\partial y^n\over\partial\sigma} \delta^3 (\vec x - \vec y) \delta (y_0 - t)$$
$$	= e \int_{\sigma_{2}}^{\sigma_{1}} \left({\partial \vec
y\over\partial\sigma} \times {\partial\vec y\over\partial t}\right)_k \delta^3
(\vec x - \vec y) = M_{k}. \eqno (2.9a)$$
Next set $\mu = i$ and $\nu = j$ in eq. (2.8):
$$	G_{ij}^s = - e \epsilon_{ij0k} \int_{\tau_{1}}^{\tau_{2}} d\tau
\int_{\sigma_{2}}^{\sigma_{1}} {\partial y^0\over\partial\tau} {\partial
y^k\over\partial\sigma} \delta^3(\vec x - \vec y) \delta (y_0 - t)$$
$$	= - e \epsilon_{ijk} \int_{\sigma_{2}}^{\sigma_{1}} d\sigma {\partial
y^k\over \partial\sigma} \delta^3 (\vec x - \vec y) = - \epsilon_{ijk} P_{k}.
\eqno (2.9b)$$
Thus eq. (2.8) is just the covariant version of eqs. (2.3) and (2.5).

Eq. (1.12) with $\mu = 1$, and $G_{\mu\nu}^s$ given by eq. (2.8) is the
covariant form of eqs. (1.14) and (1.15) determining the dual potential
produced by a pair of oppositely charged particles moving in the vacuum.  The
resulting $C_\mu$ will depend upon the location of the string, but this
dependence will drop out in the expression for the electromagnetic field tensor
$G_{\mu\nu}$.  We will show in the next section how eqs. (1.14) and (1.15) with
$\vec P$ and $\vec M$ given by eqs. (2.3) and (2.5) produce the usual
expressions for the electric and magnetic fields of slowly moving particles.

To conclude this section we note that the equation of motion (1.12) with $\mu =
1$, namely
$$	\partial^\mu G_{\mu\nu} = 0, \eqno (2.10)$$
can be obtained from a Lagrangian density ${\cal L}$ given by
$$	{\cal L} = - {1\over 4} G_{\mu\nu} G^{\mu\nu} = {1\over 2} (\vec H^2 - \vec
D^2). \eqno (2.11)$$
We will see in Section V that this Lagrangian gives not only the field
equations  but also the particle equations of motion.

\noindent {\bf III.  Coulomb's Law and the Biot Savart Law}

To understand better how dual potentials work we will solve eqs. (1.14) and
(1.15) for slowly moving particles.
First consider charges at rest.  Then $\vec M = 0$ and $\vec P$ is time
independent, and eq. (1.14) becomes $\nabla^2 C_0 = 0$, (i.e., $C_0 = 0)$, and
eq. (1.15) reduces to
$$	\vec \nabla \times (- \vec\nabla \times \vec C_D) =  \vec\nabla \times \vec
P, \eqno (3.1)$$
where we have denoted the static solution $\vec C = \vec C_D$ (for Dirac).
Eq. (3.1) has the form of the equation for the vector potential due to a
polarization current produced by the superposition (2.3) of dipoles.  Thus,
$\vec C_D$ is just the vector potential produced by a superposition of point
dipoles of strength $-ed\vec y$ distributed uniformly along the string (see eq.
2.3), i.e., $\vec C_D$ is given by$^{[5]}$
$$	\vec C_D (\vec x)  = - {e\over 4\pi} \int_{\vec x_{2}}^{\vec x_{1}} d\vec y
\times {(\vec x - \vec y)\over |\vec x - \vec y|^3} .\eqno (3.2)$$
Then
$$	-\vec\nabla \times \vec C_D = {e\over 4\pi} \left\{ \int_{\vec x_{2}} ^{\vec
x_{1}} d\vec y \vec\nabla \cdot  {(\vec x - \vec y)\over |\vec x - \vec y|^3} -
\int_{\vec x_{2}}^{\vec x_{1}} \vec d y \cdot \vec\nabla {(\vec x - \vec
y)\over |x - \vec y|^3}\right\}$$
$$	= {e\over 4\pi} \left\{ \int_{\vec x_{2}}^{\vec x_{1}} d \vec y 4 \pi \delta
(\vec x - \vec y) + {(\vec x - \vec x_1)\over |\vec x - \vec x_1|^3} - {(\vec x
- \vec x_2)\over |\vec x - \vec x_2|^3}\right\}$$
$$	= \vec P + \vec D_C , \eqno (3.3)$$
where
$$	\vec D_C  \equiv \vec D_{Coulomb} = {e\over 4\pi} \left({(\vec x - \vec
x_1)\over |x - \vec x_1|^3} - {(\vec x - \vec x_2)\over |\vec x - \vec
x_2|^3}\right), \eqno (3.4)$$
so that
$$	\vec D = - \vec\nabla \times \vec C_D - \vec P = \vec D_C. \eqno (3.5)$$
The above elementary derivation of Coulomb's law indicates that it really isn't
too much harder to work with dual potentials and strings than to work with
ordinary potentials and localized charges.  The string cancellation mechanism
in eq. (3.5) is depicted in Fig. 3 in which we have taken the string to be a
straight line connecting $\vec x_2$ and $\vec x_1$.
We see that $-\vec \nabla \times \vec C_D$ gives a divergence free field
distribution.  The singular field passing through the line $L$ is cancelled by
the singular polarization $\vec P$, leaving a Coulomb field with a source at
$\vec x_1$ and a sink at $\vec x_2$.

Next, let us solve eqs. (1.14) and (1.15) to first order in $\vec v_1$ and
$\vec v_2$  and to zero order in the accelerations ${\dot{\vec v}}_1$ and
${\dot{\vec v}}_2$.  First look at eq. (1.15).  We choose the gauge  $\vec
\nabla \cdot \vec C = 0$.  (Note $\vec\nabla \cdot \vec C_D = 0$).  Then eq.
(1.14) becomes
$$	- \nabla^2 C_{0D} =  - \vec\nabla \cdot \vec M, \eqno (3.6)$$
where we have denoted the solution $C_0 = C_{0D}$.  Eq. (3.6) has the form of
the equation for the scalar potential due to a polarization charge produced by
the superposition (2.5) of dipoles.  Hence $C_{0D}$ is just the scalar
potential produced by a superposition of point dipoles of strength $e d\vec y
\times {\dot{\vec y}}$  distributed uniformly along the string, i.e., $C_{0D}$
is$^{[6]}$
$$	C_{0D} = {e\over 4\pi} \int_{\vec x_{2}(t)}^{\vec x_{1}(t)} (d\vec y \times
{\dot{\vec y}}) \cdot {(\vec x - \vec y)\over |\vec x - \vec y|^3}. \eqno
(3.7)$$
{}From eqs. (2.5) and (3.7) we see that the time derivative of $\vec M$ and
$\vec C_{0D}$ do not contain terms linear in the velocities.  The same is true
for $\partial_0^2 \vec C_D$ calculated from eq. (3.2) with $\vec x_1
\rightarrow    \vec x_1(t), \vec x_2 \rightarrow \vec x_2(t)$.  Hence to first
order in the velocities eq. (1.15) reduces to eq. (3.1), and so $\vec C = \vec
C_D$ and $\vec D = \vec D_C$.

To calculate $\vec H$ we use eq. (1.9) with $\vec C = \vec C_D$ and $C_0 =
C_{0D}$.  We first calculate
$$	- {\partial\over\partial t} \vec C_D = {e\over 4\pi} ~
{\partial\over\partial t} \int_{\vec x_{2}(t)}^{\vec x_{1}(t)} d\vec y \times
{(\vec x - \vec y)\over |\vec x - \vec y|^3}. \eqno (3.8)$$
The evaluation of the right-hand side of eq. (3.8) parallels that of eq. (2.6)
and we obtain
$$	- {\partial\over\partial t} \vec C_D = \vec H_{BS} + {e\over \delta t} \oint
d\vec y \times {(\vec x - \vec y)\over |\vec x - \vec y|^3}, \eqno (3.9)$$
where the term
$$	\vec H_{BS} \equiv \vec H_{Biot~ Savart} \equiv {e\over 4\pi} \left\{\vec
v_1 \times {(\vec x - \vec x_1 (t))\over |\vec x - \vec x_1(t)|^3}  - \vec v_2
\times {(\vec x - \vec x_2(t))\over |\vec x - \vec x_2(t)|^3}\right\}, \eqno
(3.10)$$
arises from time differentiation of $\vec x_1(t)$ and $\vec x_2(t)$ in eq.
(3.8) leaving the path fixed.  The line integral in (3.9), over the same
contour occuring in eq. (2.6), arises from moving the string keeping the
endpoints fixed.  Paralleling  eq. (2.7) we then apply Stokes' theorem to
obtain
$$	{e\over\delta t} \oint {d\vec y \times\over 4\pi} {(\vec x - \vec y)\over
|\vec x - \vec y|^3} = - {e\over\delta t} \int {(d\vec S \times
\vec\nabla_y)\over 4\pi} \times {(\vec x - \vec y)\over |\vec x - \vec y|^3}$$
$$	= e \int_{x_{2}(t)}^{\vec x_{1}(t)} \left(d \vec y \times {\delta\vec
y\over\delta t}\right) \left(\vec \nabla_y \cdot {(\vec x - \vec y)\over |\vec
x - \vec y|^3}\right)$$
$$	+\vec \nabla_x {e\over 4\pi} \int d\vec y \times {\delta \vec y\over\delta
t} \cdot {(\vec x - \vec y)\over |\vec x - \vec y|^3}$$
$$	= - \vec M + \vec\nabla C_{0D}. \eqno (3.11)$$
Eqs. (3.9) and (3.11) then yield
$$	\vec H = - {\partial \vec C_D\over\partial t} - \vec \nabla C_{0D} + M =
\vec H_{BS}. \eqno (3.12)$$
Thus we see that the Biot Savart magnetic field $\vec H_{BS}$ comes from the
time derivative of the limits $\vec x_2(t)$ and $\vec x_1(t)$ in the integral
for $\vec C_D$, eq. (3.8).  The remaining string dependent part of ${\partial
\vec C_D\over\partial t}$ cancels the contribution to $\vec H$ coming from
$C_{0D}$ and $\vec M$.

We now use the solutions for $\vec C_D, \vec C_{0D}, \vec D_C$ and $\vec
H_{BS}$  to eliminate the fields in the Lagrangian $L$ given by
$$	L =  \int d\vec x {\cal L} = {1\over 2} \int d\vec x (\vec H^2 - \vec D^2),
\eqno (3.13)$$
to second order in the velocities of the charged particles.   The Lagrangian
$L$, defined as the integral over the Lagrangian density (2.11) then becomes a
function $L = L (\vec x_1, \vec x_2, \vec v_1, \vec v_2)$ only of the positions
and velocities of the charged particles.  To higher order in the velocities one
cannot eliminate the field degrees of freedom in $L$ because of the presence of
radiation.

For particles at rest we have $\vec C = \vec C_D, C_0 = 0, \vec H = 0, \vec D =
\vec D_C$ and
$$	L(\vec v_1 = \vec v_2 = 0) = - \int d \vec x {1\over 2} \vec D_C^2 =
{-e^2\over 4\pi |\vec  x_1 - \vec x_2|}, \eqno (3.14)$$
where the self-energy has been subtracted.  To first order in the velocities,
$\vec C = \vec C_D$ and $ \vec D = \vec D_C$ given by eq. (3.4) with $\vec x_1
\rightarrow \vec x_1(t), \vec x_2 \rightarrow \vec x_2(t)$.  In other words the
static field configuration follows adiabatically the motion of the charged
particles.  Furthermore since the Lagrangian $L$ is stationary about static
solutions of the field equations we have
$$	- \int d\vec x {1\over 2} \vec D^2 = - {e^2\over 4\pi |\vec x_1(t) - \vec
x_2(t)|},$$
valid to second order in the velocities $\vec  v_1$ and $\vec v_2$.

All the velocity dependence in $L$ then comes from $\int \vec H^2$ which to
second order in the velocities is
$$	{1\over 2} \int d \vec x \vec H^2 = {1\over 2} \int d \vec x (\vec H_{BS})^2
= - {1\over 2} ~ {e^2\over 4\pi R} \left[\vec v_1 \cdot \vec v_2 + {\vec v_1
\cdot \vec R \vec v_2 \cdot \vec R\over R^2}\right], \eqno (3.15)$$
where the self-energies have again been subtracted out and where $\vec R = \vec
x_1 (t) - \vec x_2(t)$.  Hence, we obtain for the second order Lagrangian $L$
(first obtained by Darwin$^{[7]}$)
$$	L (\vec x_1, \vec x_2; \vec v_1, \vec v_2) = {1\over 2} \int d\vec x [\vec
H_{BS}^2 - \vec D_C^2]$$
$$	= {-e^2\over 4\pi R} - {1\over 2} ~ {e^2\over 4\pi R} \left[\vec v_1 \cdot
\vec v_2 + {\vec v_1 \cdot \vec R \vec v_2 \cdot \vec R\over R^2}\right]. \eqno
(3.16)$$

As a final remark we connect the notation of this paper to that used in our
previous work$^{[2]}$ on QCD where we have introduced string fields $\vec D_s$
and $\vec H_s$ defined as
$$	\vec D_s \equiv - \vec P, \quad \vec H_s \equiv  \vec M,  \eqno (3.17)$$
so that eqs. (1.9) take the form
$$	\vec D = - \vec \nabla \times \vec C + \vec D_s, \quad \vec H = - \vec
\nabla C_0 - {\partial \vec C\over \partial t} + \vec H_s. \eqno (3.18)$$
The fields $\vec D_s$ and $\vec H_s$ then cancel the string contributions to $-
\vec \nabla \times \vec C$ and $- \vec \nabla C_0 - {\partial \vec C\over
\partial t}$ yielding fields $\vec D$ and $\vec H$ free of string
singularities.  For slowly moving particles this mechanism is explicitly
exhibited by eqs. (3.3) and (3.11).

\noindent  {\bf IV.   The Dual Lagrangian and the Equations of Motion of the
String}$^{[8]}$

The action $S$ describing the electromagnetic interactions of a particle of
charge $e$ and mass $m_1$ with a particle of charge $-e$ and mass $m_2$ is
$$	S = - m_1\int_{t_{2}}^{t_{1}} \sqrt{1 - \vec v_1^2} dt - m_2
\int_{t_{2}}^{t_{1}} \sqrt{1 - \vec v_2^2} dt + \int d^4 x {\cal L}, \eqno
(4.1)$$
where ${\cal L}$ is given by eqs. (2.11) and (1.10).   Varying $C_\mu$ in the
action $S$  gives the field equation (2.10).  To obtain the equations of motion
for the particles and for the string which connects them we vary the string
coordinates: $y^\lambda \rightarrow y^\lambda + \delta y^\lambda$ and
correspondingly vary the particle positions $x_1^\lambda \rightarrow
x_1^\lambda + \delta x_1^\lambda, x_2^\lambda \rightarrow x_2^\lambda + \delta
x_2^\lambda$ such that
$$	\delta x_1^\lambda (\tau)  = \delta y^\lambda(\sigma_1, \tau)$$
$$	\delta x_2^\lambda (\tau) = \delta y^\lambda(\sigma_2, \tau). \eqno (4.2)$$
Denote
$$	{\partial y^\lambda\over\partial\sigma} = y^{\prime\lambda}, {\partial
y^\lambda\over\partial \tau} = \dot y^\lambda. \eqno (4.3)$$
Then
$$	\delta \left( - {1\over 4} \int d^4 xG^{\mu\nu} G_{\mu\nu}\right) = -
{1\over 2} \int d^4 x G^{\mu\nu} {(x)} \delta G_{\mu\nu}^s (x), \eqno (4.4)$$
where
$$	\delta G_{\mu\nu}^s = - e\epsilon_{\mu\nu\lambda\sigma}
\int^{\tau_{1}}_{\tau_{2}} d\tau \int^{\sigma_{1}}_{\sigma_{2}} d\sigma
\{[\delta  \dot y^{\prime\lambda}  y^{\prime\alpha} + \dot y^\lambda \delta
y^{\prime\alpha}] \delta (x - y) + \dot y^\lambda y^{\prime\alpha}
\partial_{y\beta} \delta(x - y) \delta y^\beta\}. \eqno (4.5)$$
Hence,
$$	- {1\over 2} \int dx  G^{\mu\nu} \delta G_{\mu\nu}^s =  {e\over 2}
\epsilon_{\mu\nu\lambda\alpha} \int^{\tau_{1}}_{\tau_{2}}  d\tau
\int^{\sigma_{1}}_{\sigma_{2}} d\sigma G^{\mu\nu} (y) (\delta \dot y^\lambda
y^{\prime\alpha} + \dot y^\lambda \delta y^{\prime\alpha}) + \dot y^\lambda
y^{\prime\alpha} \partial_\beta G^{\mu\nu} (y) \delta y^\beta$$
$$	= {e\over 2} \epsilon_{\mu\nu\lambda\alpha} \int_{\tau_{2}}^{\tau_{1}} d\tau
\int_{\sigma_{2}}^{\sigma_{1}}  d\sigma \Bigg\{ - \delta y^\lambda {d\over
d\tau} (G^{\mu\nu} y^{\prime\alpha}) - \delta y^\alpha {d\over d\sigma}
(G^{\mu\nu} \dot y^\lambda)$$
$$	+ \delta y^\beta \dot y^\lambda y^{\prime\alpha} \partial_\beta G^{\mu\nu} +
{d\over d\tau} (\delta y^\lambda G^{\mu\nu} y^{\prime\alpha}) + {d\over
d\sigma} (\delta y^\alpha G^{\mu\nu} \dot y^\lambda)\Bigg\}$$
$$	= e {\epsilon_{\mu\nu\lambda\alpha}\over 2} \int_{\tau_{2}}^{\tau_{1}} d\tau
\int_{\sigma_{2}}^{\sigma_{1}} d\sigma \Bigg\{ - \delta y^\lambda
y^{\prime\alpha} \dot y^\beta \partial_\beta G^{\mu\nu} - \delta y^\alpha \dot
y^\lambda y^{\prime\beta} \partial_\beta G^{\mu\nu}$$
$$	+ \delta y^\beta \dot y^\lambda y^{\prime\alpha} \partial_\beta G^{\mu\nu} +
{d\over d\sigma} (\delta y^\alpha G^{\mu\nu} \dot y^\lambda)\Bigg\}$$
$$	= {e\over 2} \int_{\tau_{2}}^{\tau_{1}} d\tau \int_{\sigma_{2}}^{\sigma_{1}}
d\sigma \Bigg\{  (\delta y^\lambda y^{\prime\alpha} \dot y^\beta) (-
\epsilon_{\mu\nu\lambda\alpha} \partial_\beta G^{\mu\nu}-
\epsilon_{\mu\nu\beta\lambda} \partial_\alpha G^{\mu\nu}$$
$$	+ \epsilon_{\mu\nu\beta\alpha} \partial_\lambda G^{\mu\nu}) +
\epsilon_{\mu\nu\lambda\alpha} {d\over d\sigma} (\delta y^\alpha G^{\mu\nu}
\dot y^\lambda)\Bigg\}, \eqno (4.6)$$
where we have used the fact that the variations of $\delta y^\lambda$ vanish at
$\tau_2$ and $\tau_1$.

Next we use the identity
$$	- (\epsilon_{\mu\nu\lambda\alpha} \partial_\beta +
\epsilon_{\mu\nu\beta\lambda} \partial_\alpha + \epsilon_{\mu\nu\alpha\beta}
\partial_\lambda) G^{\mu\nu} = 2 \epsilon_{\mu\lambda\alpha\beta} \partial_\nu
G^{\nu\mu}\eqno (4.7)$$
and obtain
$$	- {1\over 2} \int dx G^{\mu\nu} \delta G_{\mu\nu}^s = e
\int_{\tau_{2}}^{\tau_{1}} d\tau \int_{\sigma_{2}}^{\sigma_{1}} d\sigma \delta
y^\lambda  y^{\prime\alpha} \dot y^\beta \epsilon_{\mu\lambda\alpha\beta}
\partial_\nu G^{\nu\mu}$$
$$	+ {e\over 2} \epsilon_{\mu\nu\lambda\alpha} \int_{\tau_{2}}^{\tau_{1}} d\tau
(\delta x_1^\alpha G^{\mu\nu} (x_1) \dot x_1^\lambda - \delta x_2^\alpha
G^{\mu\nu} {(x_2)} \dot x_2^\lambda). \eqno (4.8)$$
We must add to the above variation that of the particle action $S_P$:
$$	\delta S_P = \delta \int_{\tau_{2}}^{\tau_{1}} d\tau \left[- m_1 \sqrt{\dot
x_1^2} - m_2 \sqrt{\dot x_2^2}\right]$$
$$	= \int_{\tau_{2}}^{\tau_{1}} d\tau (- m_1 \dot x_{1\alpha} \delta \dot
x_1^\alpha - m_2 \dot x_{2\alpha} \delta \dot x_2^\alpha)$$
$$	= \int_{\tau_{2}}^{\tau_{1}} (m_1 \ddot x_{1\alpha} \delta x_1^\alpha + m_2
\ddot x_{2\alpha} \delta x_2^\alpha). \eqno (4.9)$$
The total change in the action $\delta S$ due to a change in particle
coordinates is
$$	\delta S = \delta S_P + \delta \left( - {1\over 4} \int dx G_{\mu\nu}
G^{\mu\nu}\right) =$$
$$	\int_{\tau_{2}}^{\tau_{1}} d\tau \Bigg[\delta x_1^\alpha \left(m_1 \ddot
x_{1\alpha} + {e\over 2} \epsilon_{\mu\nu \lambda\alpha} G^{\mu\nu} (x_1) \dot
x_1^\lambda\right)$$
$$	+ \delta x_2^\alpha \left(m_2 \ddot x_{2\alpha} - {e\over 2}
\epsilon_{\mu\nu\lambda\alpha} G^{\mu\nu} (x_2) \dot x_2^\lambda\right)\bigg].
\eqno (4.10)$$
In proceeding from (4.8) to (4.10) we used the field equation (2.10) which
eliminates the string contribution to the variation of the action in eq. (4.8).

If we had introduced further interactions$^{[2]}$ of the $C_\mu$ so that
$\partial_\mu G^{\mu\nu} \not= 0$, then there would have been additional
variations of the action arising from the first term in eq. (4.8).  In that
case Hamilton's principle $\delta S = 0$ gives
$$	e \int_{\sigma_{2}}^{\sigma_{1}} d\sigma y^{\prime\alpha} y^{\prime\beta}
\epsilon_{\mu\lambda\alpha\beta} \partial_\nu G^{\nu\mu} (y) = 0 , \eqno
(4.11)$$
in addition to the Lorentz force equations
$$	m_1 \ddot x_{1\alpha} = {e\over 2} \epsilon_{\alpha\mu\nu\lambda} G^{\mu\nu}
(x_1) \dot x_1^{\lambda}, $$
and
$$	m_2 \ddot x_{2\alpha} = {e\over 2} \epsilon_{\alpha\mu\nu\lambda} G^{\mu\nu}
(x_2) \dot x_2^{\lambda}, \eqno (4.12)$$
following from eq. (4.10).
Thus eq. (4.11) provides a boundary condition along the strings upon the
current $\partial_\nu G^{\nu\mu}$.

\noindent {\bf V.  Conclusions}

Our purpose in writing this paper is pedagogical, though of course, motivated
by our interest in using dual potentials in QCD.  We have seen how normal
classical electrodynamics can be handled completely in terms of dual
potentials, and that the use of these potentials gives the solution to the
conventional Maxwell equations for the electric and magnetic fields and leads
to the usual Lorentz force law for the motion of charged particles.  While dual
potentials provide a somewhat awkward way to solve electrodynamics when charges
are present, they are nevertheless the natural variables to describe a
dielectric medium with long distance anti-screening.  This is the underlying
reason for their utility in describing long range QCD.
\vfil\eject

\noindent {\bf References}

\item{1.}  P.A.M. Dirac, Phys. Rev., {\bf 74} (1948) 817.

\item{2.}  M. Baker, J. S. Ball and F. Zachariasen, Phys. Rev., {\bf D47}
(1993) 3021 and references given there.

\item{3.}  S. Mandelstam, Phys. Rep., {\bf 23C} (1976) 245.

\item{4.}  G. 't Hooft, in Proc. Europ. Phys. Soc. Conf. on High Energy Physics
(1975) edited by A. Zichichi (Editrice Compositori) Bologna, (1976) p.1225.

\item{5.}  D.J. Griffiths, Introduction to Electrodynamics, Prentice Hall,
(1989) p.149.

\item{6.}  Reference 5 Page 239.

\item{7.}  L. Landau and E. Lifshitz, Classical Theory of Fields, Pergamon
Press (1975) p.168.

\item{8.}  This section just repeats in somewhat more detail the corresponding
calculation in Reference 1 and is included here for completeness.

\bigskip

\noindent {\bf Figure Captions}

\item{Fig. 1}  Dirac String $L(t)$ connecting oppositely charged particles.

\item{Fig. 2}  Closed contour describing path of line integral on right-hand
side of eq. (2.6).

\item{Fig. 3}  Diagram representing string cancellation mechanism of eq. (3.5).

\bye